\documentclass[a4paper,fleqn]{cas-dc}
\usepackage{float}
\usepackage{anyfontsize}
\usepackage{expl3}
\usepackage{tabularray}
\usepackage[authoryear,longnamesfirst]{natbib}
\usepackage{enumitem}
\usepackage{makecell}
\usepackage{tabularray}
\def\tsc#1{\csdef{#1}{\textsc{\lowercase{#1}}\xspace}}
\tsc{WGM}
\tsc{QE}
\tsc{EP}
\tsc{PMS}
\tsc{BEC}
\tsc{DE}
\usepackage{booktabs}

\usepackage{array}
\usepackage{adjustbox}
\definecolor{culdesac}{HTML}{FF6B6B}
\definecolor{gridiron}{HTML}{006400}
\definecolor{organico}{HTML}{45B7D1}
\definecolor{headergray}{gray}{0.95}

\usepackage{subcaption}  
\newcommand{\subfigsuffix}{1} 

\usepackage{booktabs}
\usepackage{array}
\usepackage{longtable}
\usepackage{multirow}
\usepackage{colortbl}
\usepackage{geometry}
\usepackage{amsmath}
\usepackage{amssymb}
\usepackage{tabularray}
\UseTblrLibrary{booktabs}
\definecolor{blanco_total}{RGB}{255,255,255}
\usepackage{tabularray}
\usepackage{caption}
\usepackage{sectsty}
\usepackage{svg}
\usepackage{hyperref}
\usepackage{tikz}

\usetikzlibrary{positioning}  
\usetikzlibrary{arrows.meta} 

\sectionfont{\small\bfseries}    
\subsectionfont{\small\bfseries}       
\subsubsectionfont{\footnotesize\bfseries} 

\definecolor{culdesac_optimo}{RGB}{255, 235, 238}
\definecolor{culdesac_moderado}{RGB}{255, 205, 210}
\definecolor{culdesac_critico}{RGB}{244, 67, 54}

\definecolor{gridiron_optimo}{RGB}{232, 245, 233}
\definecolor{gridiron_bueno}{RGB}{200, 230, 201}
\definecolor{gridiron_variable}{RGB}{165, 214, 167}

\definecolor{organico_optimo}{RGB}{227, 242, 253}
\definecolor{organico_bueno}{RGB}{187, 222, 251}
\definecolor{organico_variable}{RGB}{144, 202, 249}

\definecolor{header_culdesac}{RGB}{198, 40, 40}
\definecolor{header_gridiron}{RGB}{56, 142, 60}
\definecolor{header_organico}{RGB}{25, 118, 210}
\usepackage{anyfontsize}

\begin{document}
\fontsize{8}{10}\selectfont  
\let\WriteBookmarks\relax
\def\floatpagepagefraction{1}
\def\textpagefraction{.001}

\shortauthors{Riascos-Goyes et~al.}

\title [mode=title]{Decoding street network morphologies and their correlation to travel mode choice}
\shorttitle{Urban morphology and mobility patterns}
\shortauthors{Riascos~et~al.}
\author[1]{Juan F. Riascos-Goyes\cormark[2]}[orcid=0009-0009-9599-1030,]
\ead{jfriascosg@eafit.edu.co}

\author[2]{Michael Lowry\cormark[3]}[orcid=0000-0002-6203-1502,]
\ead{mlowry@uidaho.edu}

\author[1,4]{Nicolás Guarín-Zapata\cormark[4]}[orcid=0000-0002-9435-1914,]
\ead{nguarinz@eafit.edu.co}

\author[1,3]{Juan P. Ospina\cormark[1]}[orcid=0000-0001-6374-083X,]
\ead{jospinaz@eafit.edu.co}

\affiliation[1]{organization={School of Applied Sciences and Engineering, Universidad EAFIT},
    city={Medellín},
    country={Colombia}}

\affiliation[2]{organization={Department of Civil and Environmental Engineering, University of Idaho},
    city={Moscow, ID},
    country={United States}}

\affiliation[3]{organization={Nature and Cities Research Group, Universidad EAFIT},
    city={Medellín},
    country={Colombia}}

\affiliation[4]{organization={Math Applications in Science and Engineering Research Group, Universidad EAFIT},
    city={Medellín},
    country={Colombia}}

\begin{abstract}
Urban morphology has long been recognized as a factor shaping human mobility, yet comparative and formal classifications of urban form across metropolitan areas remain limited. Building on theoretical principles of urban structure and advances in unsupervised learning, we systematically classified the built environment of nine U.S. metropolitan areas using structural indicators such as density, connectivity, and spatial configuration. The resulting morphological types were linked to mobility patterns through descriptive statistics, marginal effects estimation, and post hoc statistical testing. Here we show that distinct urban forms are systematically associated with different mobility behaviors, such as reticular morphologies being linked to significantly higher public transport use (marginal effect = 0.49) and reduced car dependence (–0.41), while organic forms are associated with increased car usage (0.44), and substantial declines in public transport (–0.47) and active mobility (–0.30). These effects are statistically robust ($p < 10^{-19}$), highlighting that the spatial configuration of urban areas plays a fundamental role in shaping transportation choices. Our findings extend previous work by offering a reproducible framework for classifying urban form and demonstrate the added value of morphological analysis in comparative urban research. The dataset and code are openly available, allowing replication and adaptation to other geographical contexts. These results suggest that urban form should be treated as a key variable in mobility planning and provide empirical support for incorporating spatial typologies into sustainable urban policy design.

\end{abstract}




\maketitle

\section{Introduction}
The physical structure of cities exerts a determining influence on the daily lives of their inhabitants, shaping the way they move, access services, and participate in urban dynamics. In this context, urban morphology stands as a crucial factor in the organization of mobility, an aspect increasingly relevant in global debates on walkability, transit-oriented development, and the 15-minute city paradigm. This morphology not only defines the accessibility and connectivity of a territory but also conditions the modal share of trips and, consequently, equitable access to the city's work, educational, and social opportunities \citep{intro1}.

Historically, urban and transport policies have tended to address mobility from a predominantly functionalist perspective, focused on the operational efficiency of transport systems. However, this approach has often underestimated how urban design can either foster or restrict certain modes of travel. While previous research has demonstrated that some urban forms promote active travel and public transit, whereas others foster car dependency \citep{intro2}, much less attention has been paid to how localized street network subpatterns shape travel behavior. This represents a critical gap in understanding how urban form influences equity and sustainability in mobility. Recent works have increasingly employed deep learning and graph embeddings to characterize urban street morphology (e.g., \cite{boeing_modeling_2024,machine_learning_for_streets_patterns, chen_classification_2021}), yet few have systematically linked these morphological clusters to behavioral outcomes such as travel mode choice. Our contribution bridges this methodological and empirical gap.

In this context, the present study examines how the morphological configuration of street networks influences urban mobility patterns. It analyzes the structural attributes of road systems as captured by topological and spatial indicators \citep{Indicadores_perspectiva_general, cardillo2006}, and specifically investigates the explanatory power of subpatterns within larger networks. The central research question is: to what extent do variations in local street network subpatterns explain differences in travel mode choice, beyond socioeconomic and land-use factors? Through a comparative approach across U.S. cities, the study provides empirical evidence that can inform more integrated urban and transport planning policies aimed at promoting more equitable, sustainable, and efficient mobility systems.

\section{Urban Morphology and Spatial Structure}
\label{sec:urban_morphology_spatial_structure}
Urban morphology is a multifaceted concept that has been interpreted in various ways across disciplines such as geography, architecture, planning, and urban design \citep{Moudon1997, marshall_streets}. Some approaches emphasize the built fabric—including plots, buildings, and blocks \citep{Conzen1960}—while others focus on the configurational logic of space and the movement it generates \citep{Hillier_Hanson_1984}.

In this study, we adopt a network-based understanding of urban morphology, where the structure and geometry of the street network are treated as a proxy for the underlying form of the city \citep{Porta2006, entropia}. We acknowledge that this is only one of many valid ways to study urban form and do not claim conceptual primacy. Rather, we recognize that form is multidimensional, and that our approach captures one of its structural expressions—specifically, the topological and geometric configuration of urban streets as they relate to patterns of movement and accessibility. This computational morphology perspective aligns with the emerging field of urban systems science, where cities are analyzed as complex adaptive networks \citep{batty_building_2012,BARTHELEMY20111}

To characterize the spatial structure of cities, we draw on the principles of urban morphology, which offer a framework for describing and measuring the physical configuration of the built environment. This approach emphasizes the role of street networks as a key component in shaping urban form, since their structure often reflects identifiable patterns such as grid-like, organic, or cul-de-sac configurations.

Through the use of quantitative indicators, it is possible to capture both the topological and geometric dimensions of these patterns. Topological variables reflect the underlying structure of connectivity and potential accessibility within the street network, while geometric variables capture spatial properties, including street length, orientation, and the configuration of intersections and blocks \citep{barrington2019, barrington2019supp, jiang2007}. Together, these dimensions provide a robust analytical basis for detecting structural regularities and interpreting the spatial organization of the urban fabric in the cases analyzed.

\subsection{Characterization of Urban Morphology: Fundamental Metrics and Dimensions}
\label{ssec:characterization_morphology}
The description of an urban network, especially when approached from graph theory, generates a wide set of topological and spatial variables. Given the considerable number of these variables and their frequent intercorrelation, their direct management can be complex. To address this, the variables are grouped into a limited number of relevant dimensions that capture key aspects of urban morphology. This conceptual grouping helps consolidate numerous individual metrics into a more manageable set of descriptors with clear interpretative value. Although four principal dimensions are proposed based on logical associations among urban properties, this structure is not rigid. Depending on the complexity and scale of the dataset (e.g., in larger or more heterogeneous cities), additional or slightly different groupings may be necessary. These groupings maintain coherence with the original variables and preserve the overall interpretative framework, allowing for flexibility and nuance. From this theoretical organization, the following main dimensions are identified:
\begin{itemize}[leftmargin=*labelsep=5pt]
\item \textbf{Connectivity:} Evaluates the degree of interconnection between nodes (intersections) and segments (streets) of the network.
\item \textbf{Geometry:} Defines the physical properties of road segments and blocks, such as the average length of streets, their sinuosity, and the regularity of block shapes.
\item \textbf{Density:} Measures the concentration of road elements per unit of area.
\item \textbf{Angular Characteristics:} Describes the angles formed by road segments at intersections.
\end{itemize}

Table \ref{tab:urban_metrics} presents the seventeen metrics organized according to the previously defined categories. These metrics provide a quantification of both topological and spatial characteristics of urban street networks. The selection prioritizes not only analytical robustness, but also the capacity of these metrics to be interpreted in morphological terms, allowing for a clearer understanding of structural differences across urban contexts.

{\fontsize{7.5}{9}\selectfont
\begin{longtblr}[
  caption = {Urban street network metrics grouped by category.},
  label = {tab:urban_metrics},
]{
  colspec = {X[1.1cm,l] X[1.3cm,l] X[2.2cm, l] X[2cm,l]},
  rowhead = 1,
  stretch = 1.2,
}
\hline
Category & Metric & Definition & Value remark \\
\hline
\makecell[tl]{Connectivity}
  & Dead End \newline Ratio 
  & Share of segments terminating in dead ends.
 
  & 0 to 1 higher means less through connectivity. \\

  & L-junction \footnotemark[1]
  & Proportion of nodes with 2 connecting streets. 
  & 0 to 1 higher values reflecting lower connectivity. \\

& T-junction \footnotemark[1]
  & Proportion of nodes with 3 connecting streets. 
  &  0 to 1 and commonly found in irregular layouts. \\

& X-junction \footnotemark[1]
  & Proportion of nodes with 4 connecting streets. 
  &  0 to 1 higher values indicate greater connectivity. \\

  & Streets per  \newline Node \footnotemark[1]
  & Average number of streets converging \newline at a node. 
  & Low (dead ends) to high (junctions) complexity. \\

  & Avg \newline Degree \footnotemark[1]
  & Average node degree, measuring number \newline of connections. 
  & 0 to 1 \\

\makecell[tl]{Geometry }
  & Circuity \footnotemark[1]
  & Average ratio of shortest-path length to Euclidean distance. 
  & From near 1 (direct) to higher values (indirect). \\

  & Avg Street \newline Length \footnotemark[1]
  & Average length of continuous streets \newline in the network. 
  & Short to long streets depending on urban form. \\

\makecell[tl]{ Density }
  & Edge \newline Density \footnotemark[1]
  & Length of edges per unit area. 
  & Low to high density depending on morphology. \\

  & Street \newline Density  \footnotemark[1]
  & Number of streets per unit area in the \newline spatial boundary. 
  & Sparse to dense street patterns observable. \\

  & Node \newline Density \footnotemark[1]
  & Share of nodes (intersections) per unit area. 
  & 0 to 1 \\

  & Intersection \newline Density \footnotemark[1]
  & Number of intersections per km² within the \newline network area. 
  & Low to high intersection frequency. \\

  & Segment \newline Density \footnotemark[1]
  & Share of street segments per km² in the area. 
  & 0 to 1 \\

\makecell[tl]{ Angular \\ Properties} 
  & Mean \newline Angle 
  & Average angle formed at intersections \newline in the network. 
  & Ranges from acute to right to obtuse angles. \\

  & Angle \newline CV 
  & Coefficient of variation of intersection angles. 
  & Low to high angular variability in layout. \\

  & Orthogonal \newline Proportion 
  & Percentage of intersections forming \newline near 90º angles. 
  & 0 to 1 \\
& Orientation \newline  Entropy \footnotemark[1]
  &Normalized measuring angular uncertainty.
  & 0 to 1 higher values indicate greater directional diversity. \\
\hline
\end{longtblr}

}
\footnotetext[1]{Internals Reference – OSMnx 1.6.0 documentation: \url{https://osmnx.readthedocs.io/en/stable/internals-reference.html\#osmnx-stats-module}.}

Beyond the dimensions addressed in this study, existing literature on urban network analysis has emphasized the role of centrality measures as part of the methods used to characterize the internal structure of street systems. These metrics arise from the topological perspective inherent to graph theory and allow for the characterization of the relative position of nodes and edges within a system, beyond their physical location or immediate connectivity.

From this perspective, centrality metrics provide an additional layer of analysis that enables the examination of how accessibility and connectivity are distributed across different street fabrics. When interpreted through a functional lens, these measures help identify not only local connection patterns but also global articulation dynamics within the network. Degree centrality \citep{degree_centrality}, by estimating the number of direct connections each node has relative to the total, offers a first approximation of its level of integration; building on this, closeness centrality \citep{CC1, cc2} broadens the scope by considering the average distances between nodes, allowing for the identification of strategic points with greater potential access to the overall system. Complementarily, betweenness centrality as elaborated by \cite{degree_centrality}, introduces a critical dimension by revealing which nodes tend to lie along the shortest paths between other pairs, highlighting their role as articulators of general flow. A similar logic can be applied to the network’s segments through edge betweenness centrality \citep{EBC}, which helps identify links with a high structural weight in overall connectivity. Although these metrics were not part of the empirical analysis developed in this study, they represent a complementary analytical dimension that can enrich the study of spatial configurations and the structural understanding of urban networks. These were excluded from the main clustering pipeline to prioritize variables with stable distributions across diverse urban contexts and to avoid multicollinearity issues in PCA.

\subsection{Typologies of Urban Patterns}
\label{ssec:typologies_urban_patterns}
The systematic analysis of the morphological metrics and dimensions described above leads to the identification and classification of different typologies of urban patterns. These represent spatial configurations that, although they may present local variations and degrees of mixing in urban reality, share distinctive structural characteristics. A clear understanding of urban typologies is fundamental for analyzing the morphology of cities. In this regard, \cite{street_standards_1995} offers a comprehensive historical overview of the evolution of street patterns. Building on this foundation, the existing literature \citep{marshall_streets, MORFOLOGIA2} has identified several predominant typologies, among which the following stand out as particularly relevant to this study:

\subsubsection{Gridiron (Reticular)}
Characterized by a predominantly orthogonal street pattern where streets intersect at right angles, generating regularly shaped blocks. This structure tends to offer high connectivity and permeability, facilitating orientation and efficient distribution.

\subsubsection{Suburban (Cul-de-sac)}
Presents a hierarchical structure, frequently with a dendritic design. It is distinguished by a high percentage of dead-end streets (\textit{cul-de-sacs}) that feed into collector roads and, finally, main arteries.

\subsubsection{Organic (or Irregular)}
Often arises from more spontaneous urban growth, adapted to topography, or historically developed without a unified geometric plan. It is defined by streets with non-uniform layouts, variable widths, and blocks of diverse shapes and sizes.

\subsubsection{Hybrid}
In reality, many urban areas do not strictly conform to a single typology but instead present a combination of characteristics from the aforementioned patterns. These hybrid patterns can arise from the superposition of different planning phases, adaptation to specific geographical contexts, or the organic evolution of previously planned areas.

\begin{figure}[htbp]
    \centering
    \includegraphics[width=\linewidth]{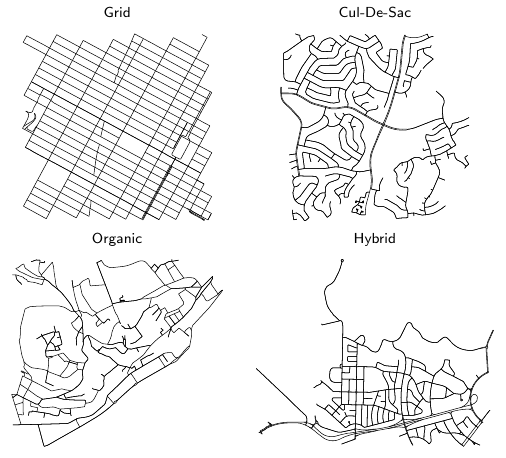}
    \caption{Illustrative cases of canonical urban street typologies. (Produced by the author using OpenStreetMap data).}
    \label{fig:urban_typologies}
\end{figure}

To illustrate the morphological diversity discussed above, Fig. \ref{fig:urban_typologies} presents representative examples of the four canonical street network patterns. These include the grid layout of Midtown Manhattan (New York City), the dendritic cul-de-sac structure of Mission Viejo (California), the organic fabric of Alfama (Lisbon), and the hybrid configuration observed in Canberra (Australia).

\section{Methodology for Theoretical Classification and Clustering of Urban Patterns}
\label{sec:methodology_theoretical_classification}

To conduct the morphological analysis of the street networks in the selected cities, the urban territory was divided into administrative units defined by the Census Bureau, known as census tracts \citep{uscensus2024}. This territorial segmentation enables a precise and consistent characterization of urban morphology, facilitating the integration of topological and spatial properties with complementary data such as mobility indicators available at the tract level. This approach enables systematic comparisons across urban areas while preserving the granularity needed to capture internal heterogeneity.

Figure~\ref{fig:methodology} presents the overall methodological framework employed in this study. The analytical process begins with the extraction of spatial and topological metrics from street networks, which are subsequently used for theoretical classification through a Multi-Attribute Decision Making (MADM) approach. In parallel, unsupervised clustering techniques combining PCA and K-means are applied to reveal emergent subpatterns within the theoretically defined typologies, enabling a comprehensive characterization of urban morphology at multiple scales.

\begin{figure*}[htbp]
\centering
\begin{tikzpicture}[
  node distance=2.2cm,
  box/.style={rectangle, draw={rgb,255:red,134;green,149;blue,159}, 
              fill={rgb,255:red,134;green,149;blue,159}, fill opacity=0.15,
              text width=3.5cm, align=center, minimum height=1.2cm, 
              rounded corners=2pt, thick, font=\small, text opacity=1},
  process/.style={rectangle, draw={rgb,255:red,115;green,106;blue,141}, 
                  fill={rgb,255:red,115;green,106;blue,141}, fill opacity=0.18,
                  text width=3.5cm, align=center, minimum height=1.2cm,
                  rounded corners=2pt, thick, font=\small, text opacity=1},
  result/.style={rectangle, draw={rgb,255:red,142;green,172;blue,176}, 
                 fill={rgb,255:red,142;green,172;blue,176}, fill opacity=0.20,
                 text width=3.5cm, align=center, minimum height=1.2cm,
                 rounded corners=2pt, thick, font=\small, text opacity=1},
  arrow/.style={->, >=stealth, thick, black!70},
  label/.style={font=\footnotesize\itshape, text=black!60}
]

\node[box] (data) {Spatial and\\Topological Metrics};

\node[process, right of=data, xshift=3cm] (madm) {Theoretical\\Classification\\(MADM)};

\node[process, right of=madm, xshift=3cm] (cluster) {Clustering\\(PCA + K-means)};

\node[result, below of=madm, yshift=-1cm] (typology) {Theoretical\\Typologies};

\node[result, below of=cluster, yshift=-1cm] (subpatterns) {Typology +\\Subpatterns};

\draw[arrow] (data) -- node[above, label] {input} (madm);
\draw[arrow] (madm) -- node[above, label] {features} (cluster);
\draw[arrow] (madm) -- node[left, label] {classification} (typology);
\draw[arrow] (cluster) -- node[right, label] {refinement} (subpatterns);

\draw[arrow, dashed, black!40] (typology) -- node[below, label] {comparison} (subpatterns);

\end{tikzpicture}
\caption{Methodological framework for typology classification and subpattern discovery. The process integrates spatial and topological metrics through theoretical classification (MADM) and unsupervised clustering to identify urban typologies and their internal subpatterns.}
\label{fig:methodology}
\end{figure*}

Following this framework, urban morphology is analyzed based on the variables previously defined and summarized in Table~\ref{tab:urban_metrics}, which group key spatial and topological characteristics relevant to describing street networks. Building on this foundation, a systematic and replicable classification method is applied to quantify the degree of correspondence between each urban unit and the theoretical profiles of urban patterns established in the literature (Section~\ref{ssec:typologies_urban_patterns}). This method employs a Multi-Attribute Decision Making (MADM) \citep{triantaphyllou2000} framework that assigns weighted scores and penalties according to how closely the observed morphological attributes align with the characteristic values of each pattern. Within this framework, three types of intervals are identified. The optimal interval includes values that are highly characteristic and representative of a specific pattern, the moderate interval comprises values compatible with the pattern but less distinctive or potentially overlapping with others, and the critical or penalizing interval encompasses values that significantly contradict the defining features of the pattern. For example, a very high proportion of dead-end streets is considered critical when evaluating the Gridiron pattern, triggering a penalty within the scoring scheme. To quantify the correspondence between an urban area and each profile, a weighted scoring mechanism with penalties is applied. Let \( d_i \) denote the value of dimension \( i \), and \( w_i \) the weight assigned based on its relevance to the pattern. The partial score for each dimension is determined by the interval in which \( d_i \) falls.
\[
s_i = 
\begin{cases}
+ S_o \times w_i, & \text{if } d_i \in \text{optimal interval} \\
+ S_m \times w_i, & \text{if } d_i \in \text{moderate interval} \\
- P_c \times w_i, & \text{if } d_i \in \text{critical interval}
\end{cases}
\]

where \( S_o \), \( S_m \), and \( P_c \) are positive coefficients that weight the contribution or penalty accordingly, with \( S_o > S_m \geq 0 \). The total score for a given urban area and a specific pattern is obtained by summing the partial scores:

\begin{equation*}
    S_{\text{total}} = \sum_{i} s_i
\end{equation*}

This approach enables a quantitative and replicable assessment of morphological similarity between observed urban areas and theoretical patterns, facilitating a systematic classification based on selected morphological properties. As previously indicated in Table~\ref{tab:urban_metrics}, this study employs a set of topological and spatial properties to characterize and classify urban morphological patterns. The definition of reference values and corresponding ranges for each metric is not arbitrary, but rather grounded in a detailed review of previous studies on urban street networks from various analytical perspectives. Regarding connectivity, multiple metrics have been explored to capture the structural integration of networks. Research such as that by \citet{machine_learning_for_streets_patterns, barrington2019} introduces indicators like the Street-Network Disconnectedness Index (SNDi), a graph-theoretic measure that quantifies structural disconnectedness using a global dataset encompassing over 46 million kilometers of streets. These metrics incorporate variables such as the proportion of dead-end streets, the continuity of links according to their hierarchical classification, and the presence of redundant connections. Complementarily, the work of \citet{jiang2007}, which analyzes networks from 40 cities in the United States and abroad, shows that urban configurations tend to exhibit properties characteristic of small-world and scale-free systems—both in terms of street length distribution and connectivity degree—thus providing a robust empirical basis for establishing reference intervals for the topological variables used. Along similar lines, the study by \citet{lowry_comparing_2014} compares 18 morphological metrics across over 500 neighborhoods in Salt Lake County, identifying which ones more effectively differentiate between historical urban development types and showing that, despite smart growth policy efforts, patterns of sprawling urbanization persist.

With respect to density, studies such as \citet{cardillo2006} and again \citet{jiang2007} have examined the concentration of street elements through weighted spatial graph representations, comparing real networks with synthetic ideal structures using methodologies such as Minimum Spanning Trees (MST) and Greedy Triangulations (GT). These approaches have demonstrated the effectiveness of these measures in capturing the structural complexity of real urban contexts. Additionally, to describe geometric and angular features of networks, studies such as \citet{orientation_e_i} and \citet{entropia} have analyzed indicators like orientation entropy, connection patterns (ringness, treeness, beltness, among others), and directional continuity. These metrics have been applied to both idealized configurations (e.g., 90°, 45°, and 30° grids) and empirical data from 100 cities across different continents, enabling the evaluation of geometric order and regularity in street orientations.

The properties selected for analysis were adapted from the ranges and threshold values reported in these studies. This information was integrated into the proposed classification framework, ensuring both conceptual consistency with the literature and empirical viability for implementation, thus enabling a structured and reproducible assessment of the correspondence between observed urban forms and the theoretical patterns considered.

\subsection{Pattern Classification and Identification of Hybrid Forms}
\label{ssec:pattern_classification_hybrid}

The classification of an urban area is determined by assigning it to the pattern typology for which it achieves the highest aggregate score, provided that this score exceeds a minimum threshold ensuring a meaningful correspondence. For example, a city is categorized as Gridiron if the score associated with that pattern is the highest among all evaluations and reaches a predefined confidence level.

Hybrid forms are identified in cases where an urban area obtains high scores in two or more distinct typologies, indicating a significant combination of characteristics from each pattern. This can be observed, for instance, in areas exhibiting prominent Gridiron traits blended with elements typical of Organic growth, often resulting from adaptations to topography or historical layering. A hybrid form is also recognized when no single score clearly dominates, yet the specific distribution of values across morphological dimensions reveals a discernible mixed configuration—for example, a predominantly orthogonal layout that includes a substantial number of dead-end streets, a feature typical of Suburban patterns in recent developments. This allows for a more nuanced classification that moves beyond mutually exclusive categories and more accurately reflects the complexity and diversity of urban fabric.

\subsection{Cluster-Based Pattern and Subpattern Classification}
\label{ssec:clasificacion_clustering_subpatrones}

The application of Principal Component Analysis (PCA) \citep{jolliffe2002} to the initial set of topological and spatial variables enabled the reduction of dimensionality, identifying a limited number of latent dimensions that capture most of the variability observed in urban morphology. To determine the appropriate number of components, we analyzed their explained variance and examined the exponential decrease in reconstruction error, adopting a $95\%$ explained variance threshold. This approach ensured a minimal loss of information while avoiding the inclusion of irrelevant components in the subsequent analysis.

Although Table~\ref{tab:urban_metrics} defined conceptual groups to organize morphological properties, the PCA analysis shows that the empirical expression of these categories may vary depending on urban scale and internal heterogeneity. Nevertheless, the extracted dimensions tend to preserve the structural logic of the theoretical classification, suggesting a robust correspondence between conceptual patterns and actual urban configurations, without implying a rigid segmentation. Based on these dimensions, clustering techniques were applied to identify predominant morphological patterns, selecting the optimal number of clusters using metrics such as the silhouette score. Although the resulting groupings reflect configurations consistent with the theoretical framework, they should be understood as flexible structures in complex urban environments such as those with large territorial extensions or multilevel developments. Peripheral subgroups may emerge that, despite deviating from the cluster centroid, retain fundamental structural properties that justify their classification.
All analyses were conducted in \texttt{Python 3.11} using the \texttt{OSMnx} and \texttt{scikit-learn} libraries. 
The complete reproducibility pipeline is available at 
\href{https://github.com/AppliedMechanics-EAFIT/urban_morphology}{the project's GitHub page}.

\section{Selection of Case Studies}

The selection of cities was guided by the principle that each case should clearly represent one of the major morphological archetypes of urban street networks: \textit{grid}, \textit{organic}, \textit{hierarchical (cul-de-sac)}, and \textit{hybrid}. This typological contrast enables both the validation of the proposed spatial and topological metrics across distinct structural regimes and the exploration of transitions between idealized and mixed forms. Each city was selected as a canonical or transitional representative of its morphological class, ensuring both historical depth and diversity of urban contexts.

\subsection{Grid Pattern: Philadelphia and Salt Lake City}

Philadelphia (Pennsylvania) represents one of the earliest and most influential grid plans in North America. Conceived by William Penn and Thomas Holme in 1682, its rectilinear layout with wide, uniform streets and five public squares reflects an intentional model for an ordered and fire-resistant city \citep{southworth2003streets,reps1965making,jackson1985crabgrass}. Conversely, Salt Lake City (Utah) illustrates a grid derived from ideological planning principles, following Joseph Smith’s \textit{Plat of Zion} and implemented by Brigham Young in 1847 \citep{greenspan2016platofzion}. The city’s grid, aligned with the cardinal axes and centered on the Temple, features unusually wide streets (approximately 40 m), designed to accommodate ox-drawn carts, thus reinforcing a symbolic and functional vision of order and expansiveness.

\subsection{Organic and Early Hybrid Patterns: Boston and Santa Fe}

Boston (Massachusetts) exemplifies an organically evolved urban morphology, shaped by topography and historical accretion rather than centralized planning \citep{warner1978streetcar,southworth1993evolving}. The street system of the Shawmut Peninsula follows preexisting paths and contours, producing a highly irregular network with variable block geometry and intersection angles. Although later interventions sought to regularize the city, most notably in the Back Bay expansion, \citep{whitehill2000boston} the influence on the historic core remains a paradigmatic organic pattern. Similarly, Santa Fe (New Mexico) combines planned and adaptive principles: founded under the Spanish Laws of the Indies with a central plaza \citep{crouch1982spanish}, its subsequent growth accommodated the terrain and existing trails, yielding a semi-regular yet curvilinear form typical of early hybrid developments.

\subsection{Hierarchical Pattern (Cul-de-Sac): Peachtree City, Cary, and Chandler}

This group encompasses suburban cities whose layouts are dominated by dendritic hierarchies and dead-end streets. Peachtree City (Georgia), established in 1959, is a model of mid-century planned suburbanism structured into self-contained villages \citep{garvin2002american}. Its defining feature is an extensive system of cul-de-sacs and loops connected by over 160 km of light-vehicle paths \citep{otoole2009gridlock}. Cary (North Carolina) similarly reflects the proliferation of cul-de-sacs through Planned Unit Developments (PUDs); however, recent urban plans encourage higher street connectivity \citep{cary2020plan}.  Chandler (Arizona), while originally agricultural and gridded, experienced massive postwar suburban expansion that replaced large blocks with residential subdivisions characterized by dendritic and cul-de-sac structures \citep{southworth2003streets}.

\subsection{Hybrid and Overlapping Patterns: Charleston and Fort Collins}

Charleston (South Carolina) and Fort Collins (Colorado) illustrate transitional and composite urban fabrics where multiple planning logics coexist. Charleston originated from the \textit{Grand Model} of 1672, introducing a colonial grid that was later modified by coastal topography and incremental development \citep{reps1965making}. The resulting fabric juxtaposes rectilinear order with organic irregularity. Fort Collins, in turn, presents a postwar superposition of morphological layers: a compact, walkable grid in the historic core, surrounded by curvilinear suburban subdivisions of the automobile era \citep{handy2003planning}.

In summary, the nine selected cities constitute a coherent comparative framework to test and validate the proposed morphological metrics. Together, they span the full spectrum of urban form, from regular, high-connectivity grids to dendritic, low-connectivity networks while also including transitional configurations that reveal the fluid boundaries between planning paradigms. This diversity allows assessing the model’s robustness across contrasting structural conditions and exploring how different spatial organizations relate to connectivity, entropy, and modal mobility patterns at the urban scale.

\section{Classification and Treatment of Urban Mobility}

In order to analyze the relationship between urban morphology and mobility patterns, data on street space usage are collected and classified based on information provided by the \cite{uscensus_mobility}. Specifically, we used ACS 5-year estimates (2017–2021) of commuting mode share at the census tract level, normalized by working-age population. This information is organized following the ABC of Mobility framework \cite{abc}, which classifies travel modes into three categories based on their function and means of transport. Active mobility includes non-motorized forms such as walking and cycling, public mobility comprises trips made using shared or collective transport systems, and private mobility refers to the use of individual motorized vehicles. This classification serves to structure the analysis of modal distribution in a clear and comparable way. Based on this classification, mobility patterns are examined from a dual perspective. On the one hand, the disaggregated approach makes it possible to assess the specific contribution of each transport mode and its potential correlation with particular morphological attributes. On the other hand, the aggregated analysis provides a general overview of the modal composition of each urban unit, allowing for systematic comparisons across different spatial contexts. This methodological approach not only enables the characterization of transport mode distribution but also facilitates the exploration of potential associations between the built environment and everyday mobility practices.

\section{Results and Comparative Analysis}

\subsection{Morphological Pattern Characterization in Selected Cities}

A sample of nine U.S. cities with diverse morphological configurations was defined, selected based on spatial variation, geographic coverage, and data availability from the United States Census. In each case, census tracts were classified according to their corresponding theoretical morphological pattern, and the resulting groupings were analyzed using clustering techniques. The aim was to explore the correspondence between the theoretical typologies and the observed configurations in each city. The results not only replicate the theoretical patterns in several cases, but also reveal the existence of subpatterns within each typology. These subgroups represent internal modulations that reflect variations within the main categories previously defined systematically in Section~\ref{ssec:characterization_morphology}. Each subgroup is labeled with a positive ($^+$) or negative ($^-$) sign, indicating an upward or downward deviation, respectively, from the dominant properties of its category.

The classification of these subpatterns enables a more precise interpretation of the intra-typological differences observed across the analyzed cities. As shown in Table~\ref{tab:subgroups}, a synthesis of these subcategories is presented, along with the general properties associated with each, fully aligned with the previously defined morphological dimensions.

{\fontsize{7.5}{9}\selectfont
\begin{longtblr}[
  caption = {Summary of morphological subgroups identified through clustering.},
  label = {tab:subgroups},
]{
  colspec = {X[2.5cm,l] X[5cm,l]},
  rowhead = 1,
  stretch = 1.2,
}
\toprule
Subgroup & General Description \\
\midrule
\texttt{Density$^\pm$} & Variations in properties related to urban density \\
\texttt{Intersection$^\pm$} & Changes in features associated with street intersections \\
\texttt{Mean$^\pm$} & Modifications in average network metrics such as segment density or circuity \\
\texttt{Street$^\pm$} & Alterations in characteristics linked to main streets and their connectivity \\
\texttt{Std$^\pm$} & Differences in geometric dispersion or variability of network angles \\
\bottomrule
\end{longtblr}
}

The cluster analysis reveals the presence of subpatterns within each general morphological category, highlighting considerable internal variability shaped by the diversity of the cities studied. Nonetheless, a strong consistency emerges, as most identified subgroups closely align with the main morphological categories previously defined. To illustrate this, Table~\ref{tab:comparacion_subpatrones} presents the nine cities analyzed, detailing their theoretical morphological classifications alongside the clustering results and the subpatterns identified.

\begin{figure*}[t]
\centering
\renewcommand{\subfigsuffix}{1}
\setcounter{subfigure}{0}
\begin{subfigure}{0.45\textwidth}
  \includegraphics[width=\linewidth]{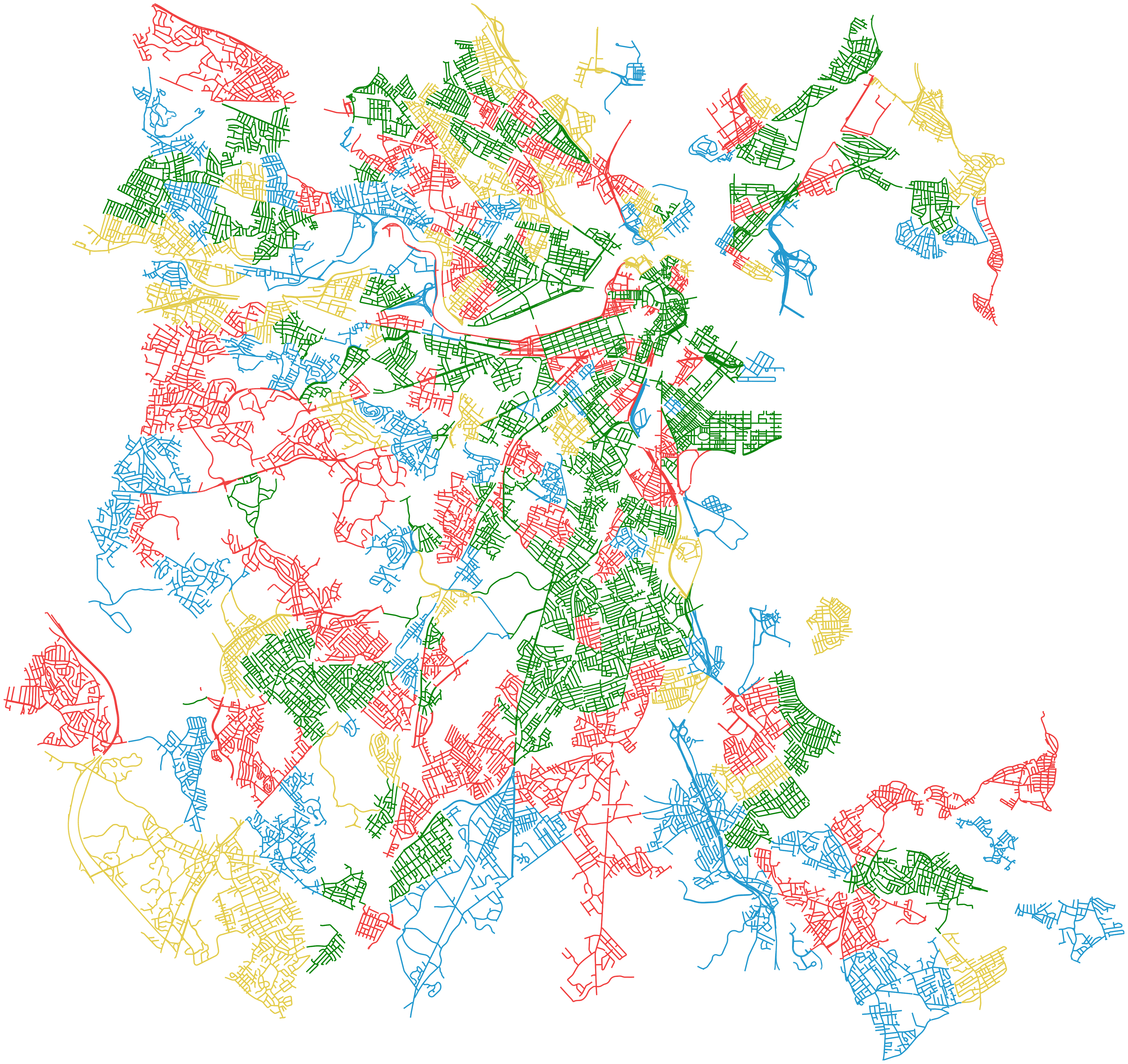}
  \caption{Boston, MA}
  \label{fig:boston_a1}
\end{subfigure}
\hfill
\renewcommand{\subfigsuffix}{2}
\setcounter{subfigure}{0}
\begin{subfigure}{0.45\textwidth}
  \includegraphics[width=\linewidth]{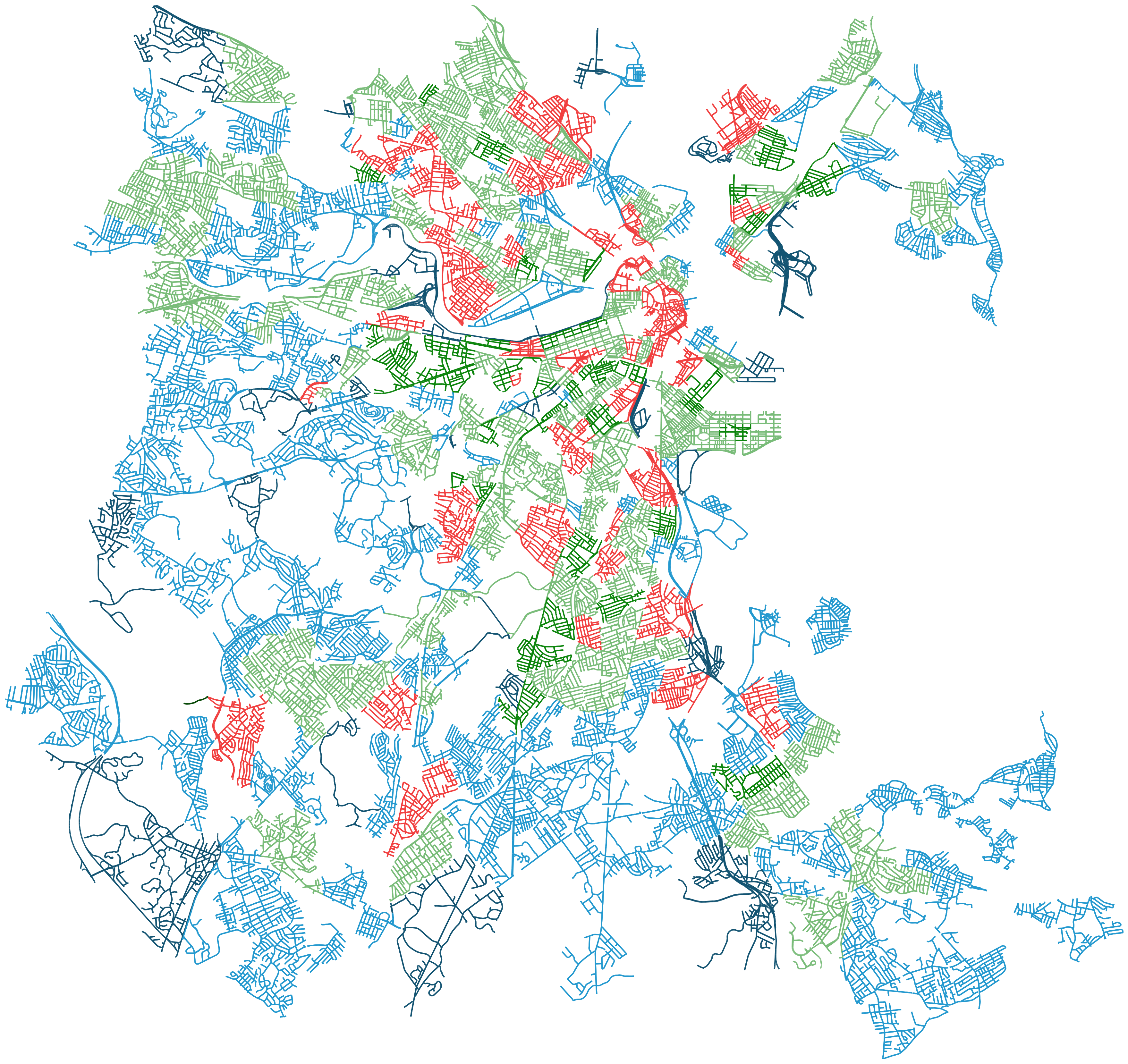}
  \caption{Boston Clustering}
  \label{fig:boston_a2}
\end{subfigure}

\vspace{1em}

\renewcommand{\subfigsuffix}{1}
\setcounter{subfigure}{1}
\begin{subfigure}{0.45\textwidth}
  \includegraphics[width=\linewidth]{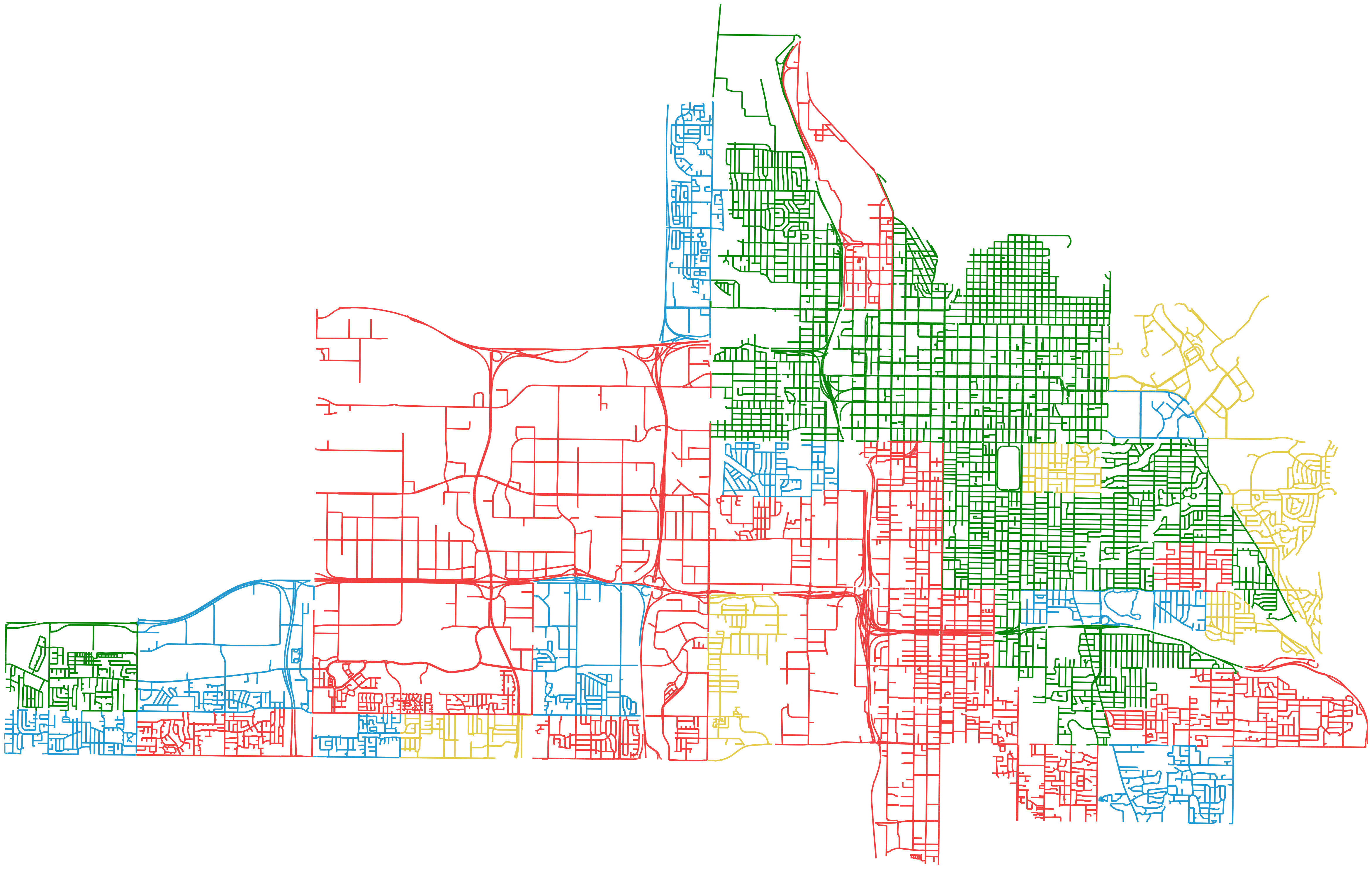}
  \caption{Salt Lake City, UT}
  \label{fig:saltlake_b1}
\end{subfigure}
\hfill
\renewcommand{\subfigsuffix}{2}
\setcounter{subfigure}{1}
\begin{subfigure}{0.45\textwidth}
  \includegraphics[width=\linewidth]{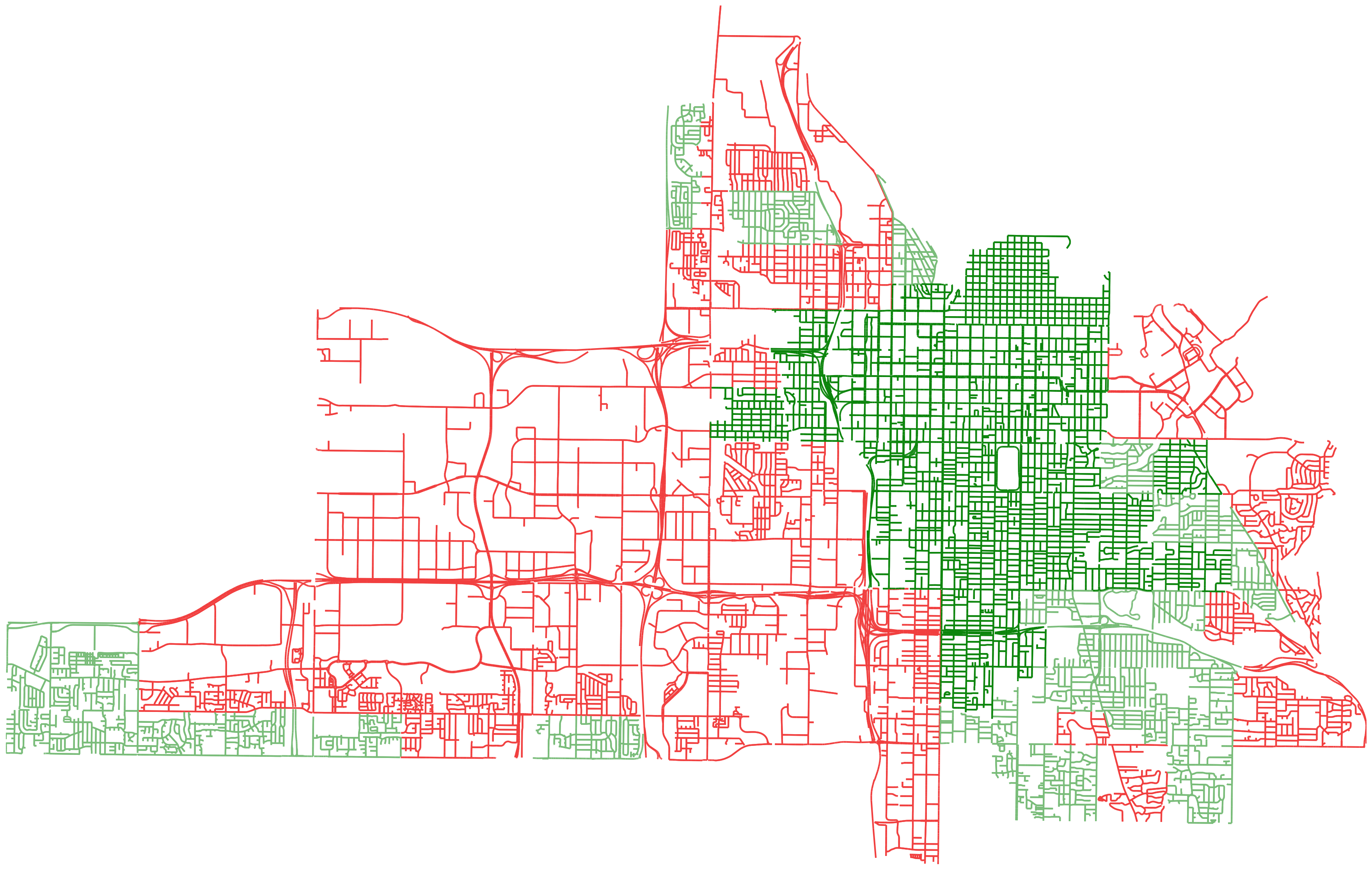}
  \caption{Salt Lake City Clustering}
  \label{fig:saltlake_b2}
\end{subfigure}

\vspace{1em}

\renewcommand{\subfigsuffix}{1}
\setcounter{subfigure}{2}
\begin{subfigure}{0.45\textwidth}
  \includegraphics[width=\linewidth]{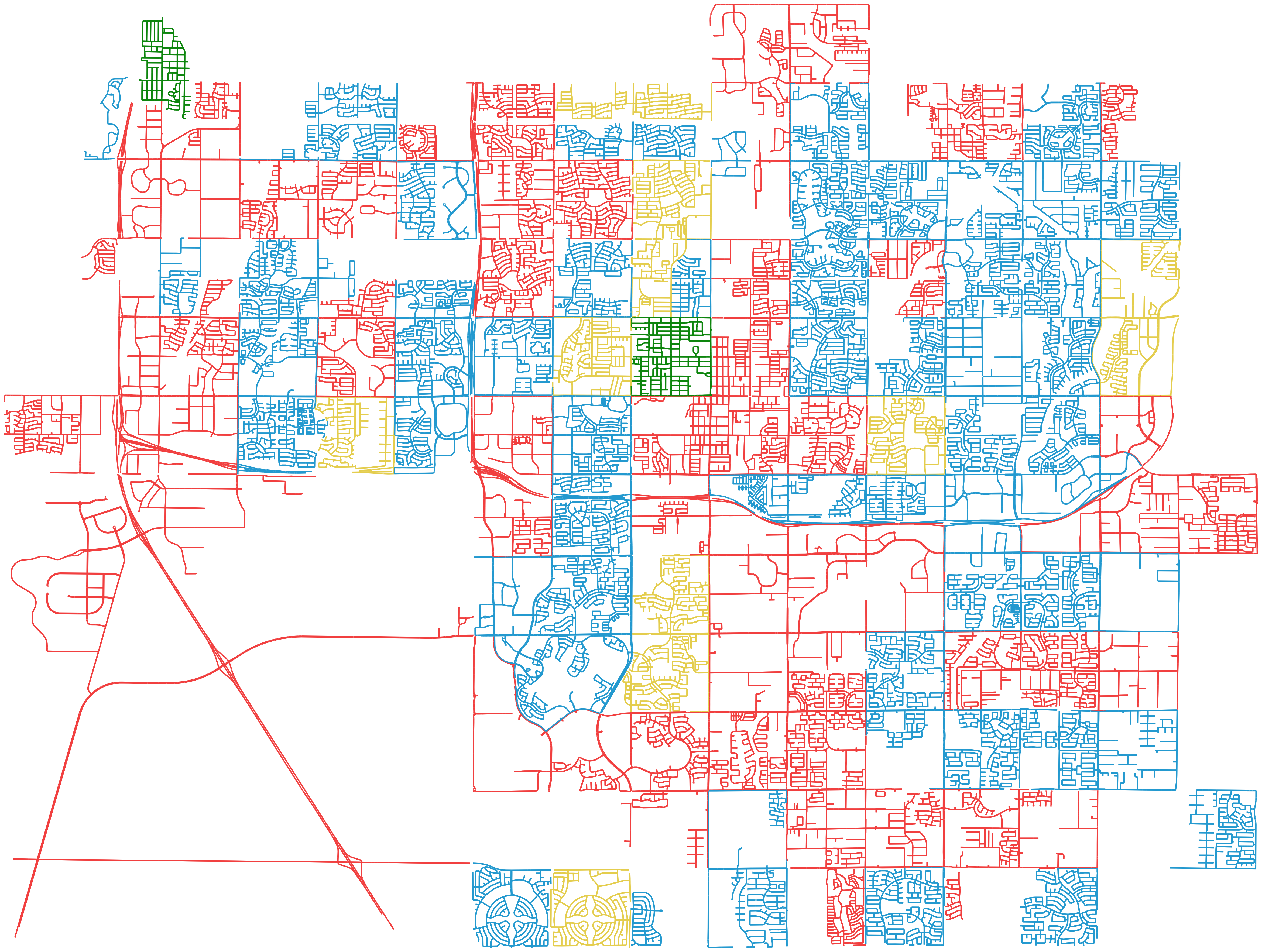}
  \caption{Chandler, AZ}
  \label{fig:chandler_c1}
\end{subfigure}
\hfill
\renewcommand{\subfigsuffix}{2}
\setcounter{subfigure}{2}
\begin{subfigure}{0.45\textwidth}
  \includegraphics[width=\linewidth]{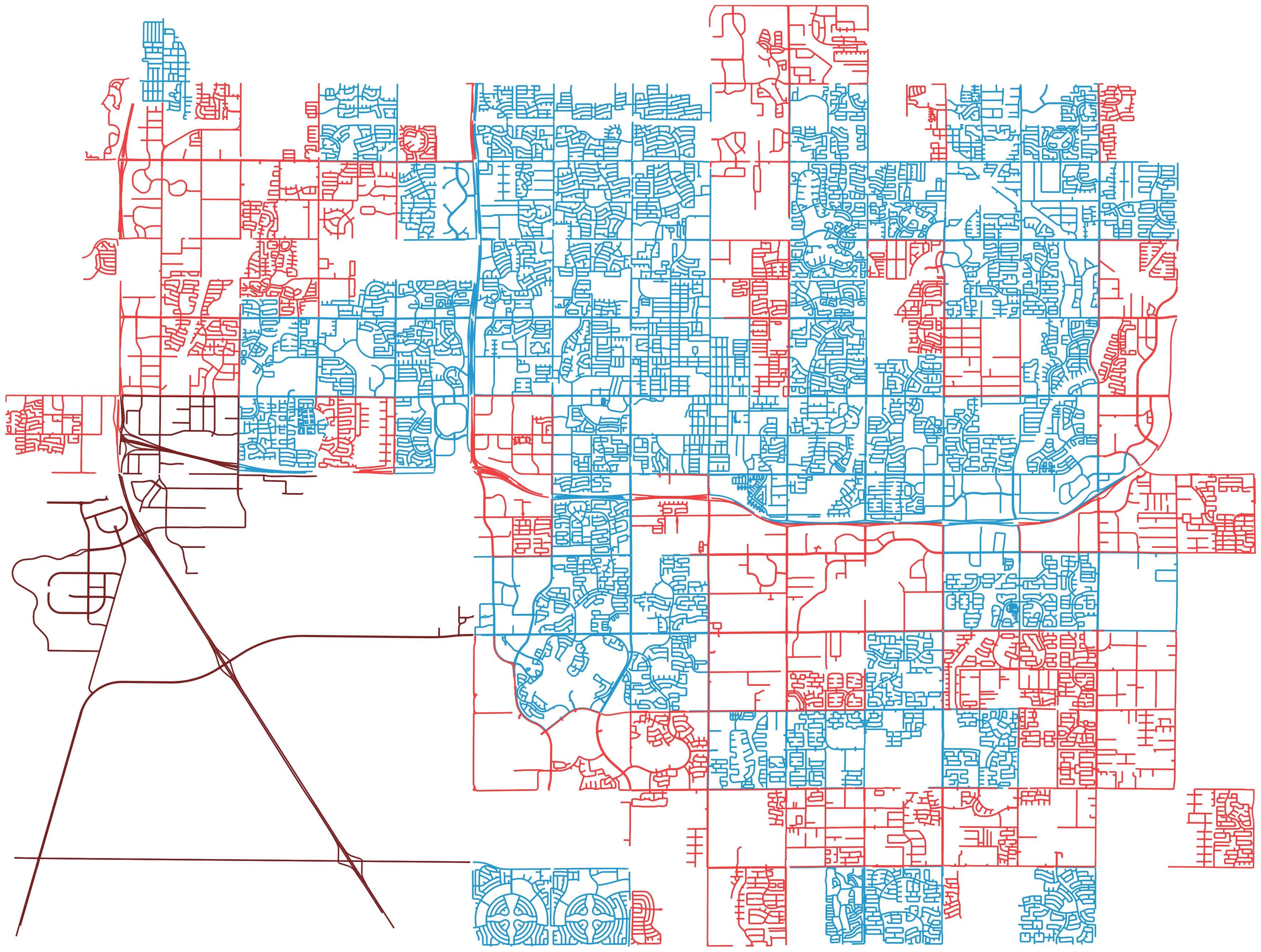}
  \caption{Chandler Clustering}
  \label{fig:chandler_c2}
\end{subfigure}

\centering
\includegraphics[width=1\textwidth]{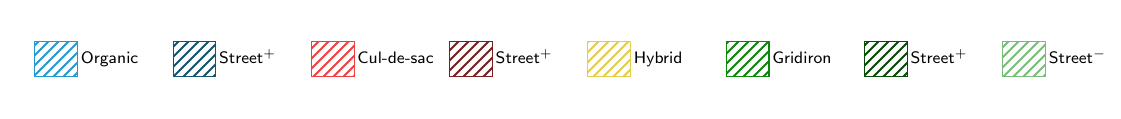}

\caption{Morphological classification of street networks. Top row (a1–c1): primary typologies. Bottom row (a2–c2): sub-patterns identified within each category via unsupervised learning.}
\label{fig:street_patterns_comparison}
\end{figure*}

{\fontsize{7.5}{9}\selectfont
\begin{longtblr}[
  caption = {Comparison between theoretical morphological patterns and sub-patterns obtained by clustering.},
  label = {tab:comparacion_subpatrones},
]{
  colspec = {X[1.2cm,l] X[2.8cm,l] X[2.8cm,l] },
  rowhead = 1,
  stretch = 1.2,
}
\hline
City & Primary Pattern & Clustering Sub-pattern  \\
\hline
\makecell[tl]{Boston}
  & Gridiron \hfill 44\% \newline Organic \hfill 18\% \newline Hybrid \hfill 14\% \newline Cul De Sac \hfill 24\% 
  & Gridiron \hfill 12.1\% \newline Street$^-$ \hfill 35.4\% \newline Street$^+$ \hfill 0.3\% \newline Organic \hfill 29.2\% \newline Street$^+$ \hfill 7.9\% \newline Cul De Sac \hfill 15.2\% \\
\makecell[tl]{Cary Town}
  & Gridiron \hfill 1\% \newline Organic \hfill 28\% \newline Hybrid \hfill 17\% \newline Cul De Sac \hfill 54\% 
  &  Gridiron \hfill 1.3\% \newline Cul De Sac \hfill 27.6\% \newline Std$^+$ \hfill 68.4\% \newline Std$^-$ \hfill 2.6\% \\
\makecell[tl]{Chandler}
  & Gridiron \hfill 2\% \newline Organic \hfill 47\% \newline Hybrid \hfill 10\% \newline Cul De Sac \hfill 41\% 
  & Organic \hfill 60.8\% \newline Cul De Sac \hfill 37.4\% \newline Street$^+$ \hfill 1.9\% \\
\makecell[tl]{Charleston}
  & Gridiron \hfill 23\% \newline Organic \hfill 39\% \newline Hybrid \hfill 8\% \newline Cul De Sac \hfill 30\% 
  &  Gridiron \hfill 11.5\% \newline Density$^-$ \hfill 9.8\% \newline Density$^+$ \hfill 4.9\% \newline Density$^-$ \hfill 11.5\% 
 \newline Organic \hfill 29.5\% \newline Density$^-$ \hfill 24.6\% \newline Cul De Sac \hfill 8.2\% \\
\makecell[tl]{Fort Collins}
  & Gridiron \hfill 12\% \newline Organic \hfill 31\% \newline Hybrid \hfill 13\% \newline Cul De Sac \hfill 44\% 
&  Gridiron \hfill 7.7\%  \newline Organic \hfill 21.2\% \newline Street$^+$ \hfill 5.8\%   \newline Cul De Sac \hfill 13.5\% \newline Street$^+$ \hfill 26.9\% \newline Street$^-$ \hfill 25.0\%\\
\makecell[tl]{Peachtree}
  &  Organic \hfill 13\% \newline Hybrid \hfill 6\% \newline Cul De Sac \hfill 81\% 
  &  Organic \hfill 12.5\% \newline Cul De Sac \hfill 18.8\% \newline Mean$^+$ \hfill 43.8\% \newline Mean$^-$ \hfill 25.0\%  \\
  
\makecell[tl]{Philadelphia}
  & Gridiron \hfill 59\% \newline Organic \hfill 16\% \newline Hybrid \hfill 11\% \newline Cul De Sac \hfill 14\% 
  & Gridiron \hfill 14.2\% \newline Mean$^-$ \hfill 41.3\% \newline Organic \hfill 44.5\% \\
\makecell[tl]{Salt Lake}
  & Gridiron \hfill 54\% \newline Organic \hfill 17\% \newline Hybrid \hfill 8\% \newline Cul De Sac \hfill 21\% 
  & Gridiron \hfill 43.1\% \newline Intersection$^-$ \hfill 25.0\% \newline Cul De Sac \hfill 31.9\% \\
\makecell[tl]{Santa Fe}
  & Gridiron \hfill 5\% \newline Organic \hfill 43\% \newline Hybrid \hfill 12\% \newline Cul De Sac \hfill 40\% 
  & Organic \hfill 26.2\% \newline Density$^-$ \hfill 19.1\% \newline Density$^+$ \hfill 2.4\% \newline Cul De Sac \hfill 19.1\% \newline Density$^+$ \hfill 16.7\% \newline Density$^-$ \hfill 16.7\% \\
\hline
\end{longtblr}
}
Figure~\ref{fig:street_patterns_comparison} presents selected examples of morphological categories and sub-patterns in cities like Salt Lake City, Boston, and Chandler, which show substantial internal variation. These cases exemplify how different urban contexts contribute to the morphological diversity captured by the clustering results, without delving into historical or social specifics.

\subsection{The Relationship Between Urban Structure and Mobility Patterns}

\begin{figure*}[!hb]
    \centering
    \includegraphics[width=\textwidth]{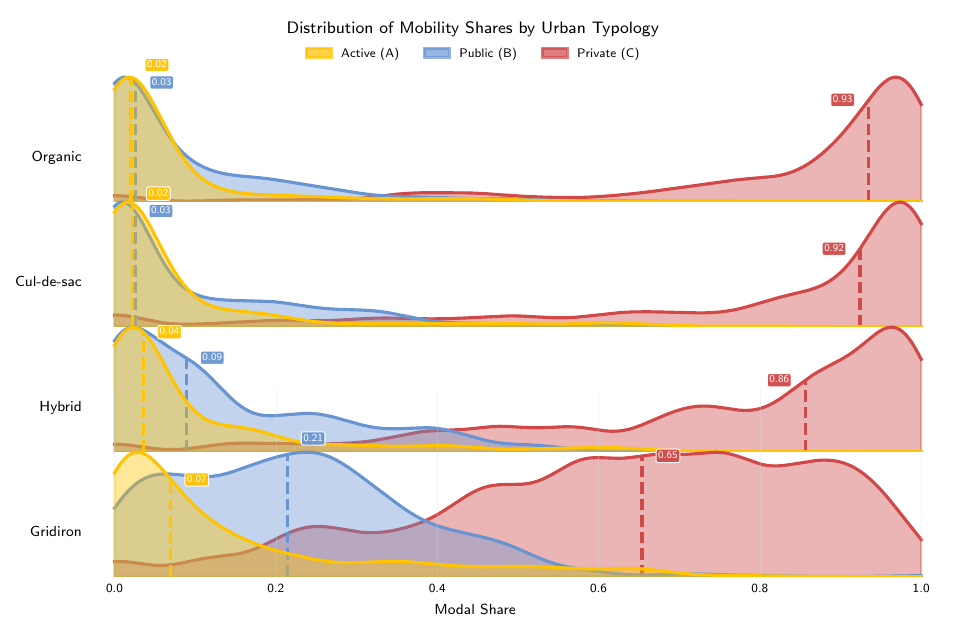}
    \caption{Ridgeline plot showing the distribution of modal shares (active, public, and private) across urban morphological patterns. Each ridge represents a density estimate for a specific typology.}
    \label{fig:density_modes_patterns}
\end{figure*}

Using data from the census on the relative shares of active, public, and private transportation modes within each urban unit, an analysis was conducted by grouping observations according to their predominant morphological pattern. This approach enabled the identification of general mobility trends associated with different spatial configurations. Figure~\ref{fig:density_modes_patterns} presents density plots illustrating how modal distributions vary across these morphological patterns.

The analysis of the median behavior and dispersion for each transportation mode reveals systematic changes that shape distinctive mobility profiles depending on urban morphology. Specifically, grid-like morphologies tend to be associated with higher shares of active and public transportation. In contrast, organic and cul-de-sac patterns—characterized by lower connectivity and a prevalence of dead-end streets—exhibit greater reliance on private vehicles. This tendency is statistically reflected in consistently lower medians for active and public transport modes within these spatial configurations.

\subsubsection{Marginal Effects of Urban Morphology on Modal Mobility}

\begin{figure}[htbp]
\centering
\includegraphics[width=\linewidth]{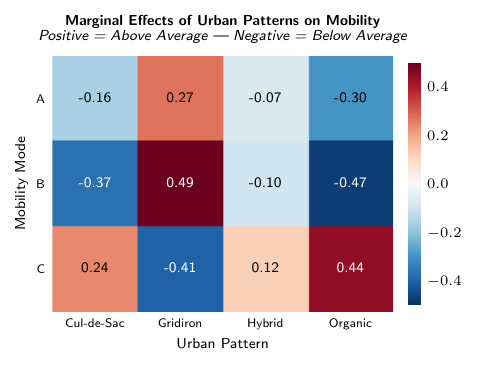}
\caption{Heatmap of the marginal effects of mobility modes across urban layout patterns.}
\label{fig:heatmap_marginal}
\end{figure}

To show how different spatial configurations are associated with variations in the use of specific modes of transportation, we analyzed the relationship between urban morphology and modal usage through estimated marginal effects. These effects quantify the deviation of each urban pattern from the overall modal share baseline, calculated as the difference between the mean modal usage within each typology and the global average across all urban units. To facilitate comparison across mobility modes with different scales, the marginal effects are normalized by the standard deviation of each variable, yielding standardized effect sizes that indicate the magnitude and direction of the association. Figure~\ref{fig:heatmap_marginal} presents a heatmap displaying these normalized coefficients for each urban pattern, allowing for a visual interpretation of how certain morphological forms are linked to positive or negative deviations in the propensity for active, public, or private mobility.
The estimated marginal effects reveal significant contrasts among the various morphological configurations. The grid-like morphology, for instance, is associated with a higher propensity for public transport use $(0.49)$ and a lower tendency toward private vehicle use $(-0.41)$, suggesting an urban environment conducive to sustainable mobility dynamics. In contrast, the organic pattern exhibits a strong positive effect on car usage $(0.44)$, along with negative effects on public transport $(-0.47)$ and active mobility $(-0.30)$, indicating a lower affinity for alternative modes of transportation in this type of urban fabric.

\subsubsection{A Post Hoc Analysis}

To more precisely assess the differences between street configurations in relation to various types of mobility, a post hoc analysis was conducted based on pairwise comparisons between patterns. The Kruskal–Wallis test \cite{kruskal1952use} was used as a global test, followed by Mann–Whitney tests \cite{mann1947test} to identify which specific contrasts were statistically significant. To estimate the magnitude of these differences, effect sizes were calculated and expressed as $\eta$. This metric allows for interpreting the practical relevance of the observed differences beyond their statistical significance. Table~\ref{tab:posthoc_mobility} summarizes the results for each combination of street pattern and mobility type, including the U statistic, associated $p$-value, and effect size $\eta$.

\begin{table*}[!hbp]
\centering
\fontsize{7.5}{9}\selectfont
\caption{Post hoc analysis results for mobility patterns across urban morphological types.}
\label{tab:posthoc_mobility}
\begin{tblr}{
  colspec = {X[1.8cm,l] X[1.5cm,l] X[1.5cm,l] X[0.8cm,l] X[0.8cm,l] X[1.5cm,l] X[1.4cm,l]  X[1.5cm,l] X[1.5cm,l] X[0.8cm,l]},
  rowhead = 1,
  stretch = 1.2,
}
\toprule
Mobility Type & Pattern 1 & Pattern 2 & n1 & n2 & U statistic & p-value & Mean Diff & Median Diff & $\eta$ \\
\midrule
\SetCell[r=5,c=1]{l} A & Gridiron & Organic & 590 & 308 & 124255 & $1.24 \times 10^{-19}$ & 0.087 & 0.049 & 0.302 \\
& Cul-de-sac & Gridiron & 344 & 590 & 69984.5 & $2.12 \times 10^{-15}$ & $-0.066$ & $-0.048$ & 0.259 \\
& Gridiron & Hybrid & 590 & 167 & 58648.5 & $1.67 \times 10^{-4}$ & 0.052 & 0.034 & 0.137 \\
& Organic & Hybrid & 308 & 167 & 20628.5 & $3.58 \times 10^{-4}$ & $-0.034$ & $-0.016$ & 0.163 \\
& Cul-de-sac & Hybrid & 344 & 167 & 24249.5 & $4.18 \times 10^{-3}$ & $-0.014$ & $-0.014$ & 0.126 \\
\midrule
\SetCell[r=5,c=1]{l} B & Gridiron & Organic & 590 & 308 & 146207.5 & $5.79 \times 10^{-51}$ & 0.143 & 0.188 & 0.501 \\
& Cul-de-sac & Gridiron & 344 & 590 & 46113 & $3.50 \times 10^{-44}$ & $-0.129$ & $-0.189$ & 0.456 \\
& Gridiron & Hybrid & 590 & 167 & 67370 & $3.92 \times 10^{-13}$ & 0.088 & 0.125 & 0.264 \\
& Organic & Hybrid & 308 & 167 & 18634 & $5.92 \times 10^{-7}$ & $-0.055$ & $-0.063$ & 0.228 \\
& Cul-de-sac & Hybrid & 344 & 167 & 22009.5 & $1.52 \times 10^{-5}$ & $-0.041$ & $-0.064$ & 0.190 \\
\midrule
\SetCell[r=5,c=1]{l} C & Gridiron & Organic & 590 & 308 & 35447 & $5.57 \times 10^{-51}$ & $-0.220$ & $-0.281$ & 0.501 \\
& Cul-de-sac & Gridiron & 344 & 590 & 150614 & $4.54 \times 10^{-35}$ & 0.168 & 0.270 & 0.404 \\
& Gridiron & Hybrid & 590 & 167 & 30866.5 & $1.65 \times 10^{-13}$ & $-0.136$ & $-0.202$ & 0.268 \\
& Organic & Hybrid & 308 & 167 & 32400.5 & $2.89 \times 10^{-6}$ & 0.085 & 0.079 & 0.215 \\
& Cul-de-sac & Hybrid & 344 & 167 & 33539 & $2.10 \times 10^{-3}$ & 0.032 & 0.068 & 0.136 \\
\bottomrule
\end{tblr}
\end{table*}

The post hoc analysis reveals a consistent relationship between urban morphology and mobility patterns, with each typology promoting or restricting different modes of travel. The gridiron configuration stands out for its sustained support of active mobility and public transport use, showing significant mean differences compared to the organic and cul-de-sac patterns—particularly in public transport mobility ($\eta = 0.50$). It also exhibits the lowest relative use of private automobiles. In contrast, cul-de-sac layout is associated with a high reliance on private vehicles and significant negative effects on active and public mobility, attributable to its limited connectivity. The mean difference of $+0.168$ compared to gridiron in private mobility ($r = 0.40$) reinforces this trend.

This tendency is even more pronounced in the organic pattern, which, due to its geometric complexity, shows the highest level of car use. The mean difference with respect to gridiron reaches $+0.220$ ($r = 0.50$), the strongest effect observed in the analysis, accompanied by the poorest performance in both active and public mobility. Finally, the hybrid morphology displays an intermediate behavior across all modes of mobility. The effect sizes relative to gridiron are smaller ($\eta$ between $0.13$ and $0.27$), suggesting that it neither optimizes any particular mode nor severely penalizes them, consistent with its mixed morphological structure.

\subsection{Policy Implications}
This study highlights the relevance of street network morphology as a key structural factor shaping mobility behavior, beyond traditional socioeconomic explanations. By revealing strong associations between street sub-patterns and mode share, the findings support the integration of morphological indicators into urban and transport policy design. Such integration can enable more spatially nuanced, evidence-based planning, with potential applications in accessibility metrics, transport equity assessments, and mobility-oriented urban development strategies, such as:
\begin{itemize}
\item \textbf{Targeted infrastructure investment for active mobility.} Cities can use morphological classifications to identify areas where the built form already supports walking and cycling. Grid-like or hybrid patterns, characterized by high connectivity and intersection density, offer fertile ground for reinforcing active mobility through additional infrastructure such as bike lanes, widened sidewalks, and traffic calming measures.
\item \textbf{Urban planning and zoning strategies.} Urban morphology can inform land-use and density policies. For instance, areas with disconnected or dendritic street patterns may require integrated planning interventions that increase permeability, street hierarchy rebalancing, or mixed-use zoning to support non-automobile travel.

\item \textbf{Spatial prioritization of public transport enhancements.} The method enables city planners to identify neighborhoods where morphological conditions hinder accessibility and multimodal integration. Such areas can be prioritized for new transit routes, improved first/last-mile connections, or flexible transport services.
\item \textbf{Scenario modeling and project impact assessments.} Morphological typologies and their modal associations can be used to simulate how changes in urban form (e.g., new developments, street reconfigurations) might affect travel behavior. This enables planners to assess the likely effectiveness of proposed projects or densification plans in shifting mode shares toward sustainable alternatives.
\item \textbf{Policy communication and participatory planning.} The visual and typological clarity of the street pattern classification offers a useful communication tool for engaging with non-technical stakeholders. It helps explain why some neighborhoods may require different strategies to foster equitable and sustainable mobility access.
\end{itemize}

\section{Discussion}

This research provides consistent evidence of a relationship between urban morphology and modal mobility patterns. As shown in Figure \ref{fig:density_modes_patterns}, environments characterized by orthogonal structures—particularly those with a gridiron pattern—exhibit a higher prevalence of active and public transportation modes compared to more fragmented layouts such as organic or cul-de-sac patterns. This tendency is functionally coherent, as grid-based layouts tend to concentrate economic, institutional, and service-related activities, thereby promoting the use of sustainable modes of transport such as walking, cycling, and public transit. Salt Lake City serves as a compelling example supporting this hypothesis which central areas such as Downtown, Central City, and Capitol Hill are consistently classified as grid-patterned in both the theoretical framework (Figure \ref{fig:saltlake_b1}) and the clustering analysis (Figure \ref{fig:saltlake_b2}), and these same areas show a modal dominance of active and public transport (Figure \ref{fig:density_modes_patterns}).

This association is further supported by quantitative results across all cities via post hoc analysis (Table \ref{tab:posthoc_mobility}). For instance, comparing gridiron and organic patterns reveals statistically significant differences in private mobility (p-value = $5.57 \times 10^{-51}$), with a negative effect size ($d = -0.220$) indicating a lower proportion of this mode in areas with a more structured street network. Similarly, the effects on active and public mobility are positive ($d = -0.281$ and $d = 0.501$, respectively), reinforcing the notion that spatial order in urban form tends to foster more sustainable transport behavior. These tendencies also appear in the comparison between cul-de-sac and gridiron patterns, highlighting the structural role of urban morphology in shaping residents' modal decisions. Overall, the findings support the central hypothesis of the study, suggesting that urban form influences not only physical accessibility but also the functional orientation of mobility systems. While this study focuses on U.S. cities, future research should expand to Global South contexts, where morphological diversity and informal development patterns may yield distinct mobility relationships.

A key limitation of this study concerns the spatial units used to segment cities, which are based on the U.S. Census Bureau’s administrative divisions, specifically the tracts. While these units effectively capture a diverse range of modal and street structure patterns, they may encompass internal heterogeneity, potentially concentrating certain mobility modes or street types and compromising the representativeness of the data. Smaller units such as blocks or sub-blocks could provide greater granularity, but their use would result in a very large number of highly homogeneous observations, increasing the likelihood of outliers and distorting aggregated indicators. For this reason, the use of tracts offers a practical balance between spatial detail and data manageability. Moreover, the analysis adopts a planar representation of urban space \citep{boeing2020planarity,bruyns}, which limits its ability to account for topographic factors such as slope, natural barriers, or elevation changes—elements that have been shown to significantly influence both mobility patterns and the structural configuration of street networks \citep{zhou2021relief,tp1}. These methodological constraints should be taken into account when interpreting the findings and point to the value of incorporating three-dimensional urban models in future research to achieve a more comprehensive understanding of spatial and functional dynamics.

Looking forward, future research should aim to deepen the theoretical and empirical understanding of how street network morphology shapes — and is shaped by — broader socio-spatial processes. While this study provided a scalable and transferable method to classify urban form and relate it to travel behavior, further work is needed to explore how these morphological patterns interact with topographic constraints, land use configurations, environmental risks, and infrastructural inequalities. In particular, we see potential in developing a critical and computational framework that bridges urban theory with topological analysis and open-source data tools. This would enable the construction of comparative morphologies across diverse geographies — especially in cities of the Global South — and support planning efforts focused on equitable, resilient, and sustainable urban transformations. By coupling computational morphology with behavioral data, this work contributes to the emerging paradigm of morphological informatics, advancing the use of data-driven typologies to inform urban mobility policy.

\section{Conclusions}

The quantitative analysis of street networks across the selected cities confirms that urban morphology constitutes a structural determinant in shaping mobility patterns. The unsupervised classification methodology not only reproduced the canonical typologies of grid, organic, and hierarchical (cul-de-sac) networks, along with their hybrid variants, but also provided a significant contribution by revealing internal variability within each category. This finer resolution demonstrates that theoretical typologies are not monolithic constructs, but rather continua that encompass measurable differences in density, connectivity, and geometric configuration, fulfilling the objective of quantifying the road network structure in detail.

A systematic and statistically significant correspondence was observed between urban form and modal split. Grid-based structures, characterized by high connectivity and permeability, consistently support higher shares of active and public transport modes. In contrast, dendritic cul-de-sac systems and irregular organic layouts are associated with a substantially greater dependence on private vehicles. The magnitude of these associations is underscored by the post hoc analyses, where the differences in automobile use between grid and organic morphologies exhibit large effect sizes, emphasizing their practical relevance. Taken together, the results reveal consistent trends linking network connectivity to transport sustainability, thereby addressing the core research objectives. Urban form thus emerges not as a passive spatial backdrop, but as an active mechanism that enables or constrains specific travel behaviors.

Overall, the findings substantiate that the configuration of the street network exerts a significant influence on urban mobility, supporting the central propositions of urban form theory. Nevertheless, the strength of these effects varies across contexts, indicating that additional factors—such as socioeconomic conditions, topography, and planning policies—mediate the relationship between morphology and mobility. This heterogeneity highlights the need for future multiscalar approaches that integrate morphological structure with broader territorial and social dynamics. Beyond validating the structural importance of urban form, this study opens avenues for comparative research in more complex and unequal urban environments, particularly in Latin American cities, where spatial fragmentation and mobility inequities pose critical challenges for sustainable urban transitions.

\bibliographystyle{cas-model2-names}

\bibliography{cas-refs}

\end{document}